\documentstyle[twocolumn,prl,aps,epsf]{revtex}
\large
\begin{document}
\epsfverbosetrue
\draft
\input epsf.sty
\twocolumn[
\hsize\textwidth\columnwidth\hsize\csname @twocolumnfalse\endcsname
\title{X-ray natural circular dichroism}
\author{Paolo Carra and Robert Benoist\cite{AAuth}} 
\address{European Synchrotron Radiation Facility,
B.P. 220, F-38043 Grenoble C\'{e}dex, France}
\date{\today} 
\maketitle 
\begin{abstract}
This paper discusses a theory of natural circular 
dichroism in the x-ray region. Integrated spectra are interpreted 
in terms of microscopic effective operators, which are derived 
in the framework of a localised (atomic) model. It is shown that the 
generators of a de 
Sitter group, such as that introduced by Goshen and Lipkin for 
nuclear structure, are suitable for describing electronic 
properties of non-centrosymmetric crystals.
\end{abstract}
]
In recent years, near-edge dichroism in crystals, 
{\it i.e}. the dependence of x-ray absorption on crystal and/or 
magnetic orientations with respect to the polarisation 
of the photon, has been thoroughly investigated at synchrotron 
radiation sources. Particular 
attention has been given to x-ray magnetic circular dichroism 
(XMCD), namely the difference in absorption between right- and left-circularly 
polarised photons in a system with a net magnetisation. Various authors 
have demonstrated the effect\cite{Sch87,Che90}, which requires 
the breaking of time-reversal symmetry and the presence of a 
spin-orbit interaction. Electric-dipole (E1) and, in some cases, 
electric-quadrupole (E2) transitions \cite{Car91,Lan95} account for the 
pertinent inner-shell excitations. 

Theoretically, efforts have been made to identify crystalline 
microscopic properties revealed by the observed spectra; and, in 
this context, atomic theory has provided a number of results. Among 
these is a set of sum rules, which relate integrated dichroic 
intensities to the ground-state expectation value of effective 
one-electron operators\cite{Tho92,Car93a,Car93b}. 
In the case of XMCD these operators coincide with the spin and orbital 
contributions to the magnetic moment, 
thus providing experimentalists 
with a simple interpretative framework. 

To cover the more realistic case of an atom in a solid, a formulation 
of the problem in terms of a minimal set of muffin-tin orbitals has also 
been reported \cite{BCA00}. In this case, corrections to the atomic results 
are found in the form of energy moments of the bands, and are expected 
to be small in most cases. 

More recently a novel phenomenon, termed x-ray natural circular 
dichroism (XNCD), was observed in an organic non-centrosymmetric 
single crystal\cite{Gou98} and in a stereogenic organometallic 
complex\cite{Ala98}. The effect stems from the interference between 
E1 and E2 transitions, thus requiring an ordered structure and the 
breaking of space inversion. 

Aimed at deriving an XNCD sum rule, 
earlier theoretical work \cite{Nat98} has identified the relevant effective 
operator with a rank-two tensor. Its microscopic nature and symmetry properties
were not, however, fully determined. 

The current paper follows the program of Ref. \onlinecite{Nat98} 
to its natural conclusion. 
Particular symmetry considerations are needed to complete the analysis; 
in virtue of them new effects, still to be probed experimentally, 
emerge. 

It is worth reminding the reader that effective operators for x-ray dichroism 
are irreducible tensors constructed from the generators of the underlying 
symmetry group. In the case of pure electric multipole-transitions, it suffices to 
work within the rotation group, {\it i.e.} with the spherical components 
of the angular momentum. In this way, sum rules for CMXD and linear x-ray
dichroism  were obtained \cite{Tho92,Car93a,Car93b}. (Linear dichroism implies 
a difference in absorption between radiations with linear polarisation 
parallel or perpendicular to a local symmetry axis.) To study interference terms  
an extension of the symmetry group is required. As shown below, the extended 
symmetry is identified with a de Sitter group, O(3,2), a non-compact version 
of O(5). Its generators will serve to write out a sum rule for XNCD. 

An aspect of symmetry relevant to XNCD is
the absence/presence of mirror planes in the permitted point groups, 
with the ensuing difference between enantiomeric and non-enantiomeric systems.
To properly elucidate this point, effective operators will be discussed for the 
E1 - magnetic dipole (M1) interference, 
which governs the natural dichroism in the optical range.

Central to our considerations is the integrated intensity
\begin{equation}
\Sigma_{\rm XNCD} = \int_{j_-+j_+}\frac{\sigma_{\rm XNCD}(\omega)}{(\hbar\omega)^2}
d(\hbar\omega)\, ,
\label{int_ncd}
\end{equation}
where $\sigma_{\rm XNCD}(\omega) = \sigma^{\bbox \epsilon}_{\rm X}(\omega) - 
\sigma^{{\bbox \epsilon}^{\ast}}_{\rm X}(\omega)$ denotes the cross 
section for natural circular dichroism in the x-ray region (X), 
a macroscopically measurable quantity.  
Integration is over a finite photon-energy range 
in Eq. (\ref{int_ncd}), corresponding 
to the two partners of a spin-orbit split inner shell. 
The partners are identified by $j_{\pm}=c\pm\frac{1}{2}$.

{\em E1-E2 interference}.  
The link between $\Sigma_{\rm XNCD}$ and a microscopic description 
is provided by the relation
\begin{eqnarray}
&&
\sigma^{\bbox \epsilon}_{\rm X}(\omega) = 4\pi^2\alpha\hbar\omega
\left[
\frac{i}{2}\sum_f 
\sum_{nn'}
\langle g\mid {\bbox \epsilon}^{\ast}\cdot {\bf r}_n
\mid f\rangle
\label{cross_section}
\right.
\\
&&
\left.
\phantom{\sum_f}
\langle f\mid {\bbox \epsilon} \cdot {\bf r}_{n'}
{\bf k}\cdot {\bf r}_{n'} \mid g\rangle 
+ {\rm c.c.}
\right]
\delta(E_f-E_g-\hbar\omega)\, ,
\nonumber
\end{eqnarray}
picking out the E1-E2 interference in the absorption cross section. 
The notation is as follows: $\hbar\omega$, ${\bf k}$ 
and ${\bbox \epsilon}$ represent energy,  wave vector and 
polarisation of the photon; $\mid g\rangle$ and $\mid f\rangle$ 
denote ground and final states of the electron system, 
with energies $E_g$ and $E_f$ respectively; electrons are labelled 
by  $n$ and $n'$; $\alpha = e^2/\hbar c$. 

We consider the fermionic field 
\begin{equation}
\Psi({\bf r}) = \sum_{l\lambda\sigma} a_{l\lambda\sigma}
\psi_{l\lambda\sigma}({\bf r}) + \sum_m a_{jm}\psi_{jm}({\bf r})\, ,
\label{ferm_field}
\end{equation}
and go over to a second quantisation description. Here, 
$a_{jm}$ and $a_{l\lambda\sigma}\;(a_{l'\lambda'\sigma'})$ 
annihilate inner-shell and valence electrons, respectively. 

The matrix element appearing in
Eq. (\ref{cross_section}) can then be given the form 
\begin{eqnarray}
&&
\sum_{nn'}
\langle g\mid {\bbox \epsilon}^{\ast}\cdot {\bf r}_n
\mid f\rangle
\langle f\mid {\bbox \epsilon} \cdot {\bf r}_{n'}
{\bf k}\cdot {\bf r}_{n'} \mid g\rangle =
\sum_{ll',jj',mm'\atop \lambda\lambda',\sigma\sigma'}
\nonumber
\\
&&
\langle\psi_{j'm'}\mid {\bbox \epsilon}^{\ast}\cdot {\bf r}
\mid\psi_{l\lambda\sigma}\rangle
\langle\psi_{l'\lambda'\sigma'}\mid {\bbox \epsilon}\cdot {\bf r}\;
{\bf k}\cdot {\bf r}\mid\psi_{jm}\rangle
\\
&&
\sum_f \langle g\mid a^{\dagger}_{jm} a_{l'\lambda'\sigma'}\mid f 
\rangle\langle f\mid a^{\dagger}_{l\lambda\sigma}a_{j'm'}
\mid g\rangle\,. 
\label{sec_quant}
\nonumber
\end{eqnarray}

The atomic basis set, which enters the definition of the fermionic 
field, is chosen as follows. Core electrons are identified by coupled 
atomic orbitals
\begin{equation}
\psi_{jm}({\bf r}) = \sum_{\gamma\sigma} 
C^{jm}_{c\gamma;\frac{1}{2}\sigma}
\varphi_{c\gamma}({\bf r})\xi_{\sigma},
\label{inner_el}
\end{equation}
with $C^{jm}_{c\gamma;\frac{1}{2}\sigma}$ 
a Clebsch-Gordan coefficient, 
$\varphi_{c\gamma}({\bf r})=\varphi_{c}
(r)Y_{c\gamma}(\hat{\bf r})$, and $\xi_{\sigma}$ a spinor.  
Valence states are described by uncoupled atomic wave functions, 
\begin{equation}
\psi_{l\lambda\sigma}({\bf r})=
\varphi_{l}(r)Y_{l\lambda}(\hat{\bf r})\xi_{\sigma}\,, 
\label{val_el}
\end{equation}
and similarly for $l'\lambda'\sigma'$. 

The derivation then proceeds by applying the Wigner-Eckart theorem 
and simple recoupling transformations. (Algebraic details are omitted 
as they can be found in the work of Natoli {\it et al.} \cite{Nat98}.) 
The result reads
\begin{eqnarray}
&&\Sigma_{\rm XNCD}=\frac{8\pi^2\alpha}{\hbar c}\sqrt{\frac{2\pi}{15}}
\sqrt{2c+1}\,\sum_{ll'}R_{cl}^{(1)}\,R_{cl'}^{(2)}
\nonumber
\\
&&
\sqrt{2l+1}\,C_{l0;10}^{c0}\,C_{c0;20}^{l'0}
\left\{
\begin{array}{ccc}
l & l' & 2 \\
2 & 1 & c\\
\end{array}
\right\}
Y_{20}(\hat{\bf k})
\label{natoli}
\\
&&
i\sum_{mm'}
\langle g| C^{20}_{lm;l'm'}a^{\dagger}_{lm}\tilde{a}_{l'm'}
- {\rm h.c. }\, |g\rangle \, ,
\nonumber
\end{eqnarray}
with the radial integrals defined by
\begin{displaymath}
R^{(L)}_{cl}=\int_0^{\infty}\!\!\!\! dr \varphi_c(r) r^{L+2} \varphi_l(r)\, .
\end{displaymath}
Also, $\tilde{a}_{lm}=(-1)^{l-m}a_{l-m}$, so that $\tilde{a}_{lm}$ and 
$a^{\dagger}_{l'm'}$ transform as the components of irreducible 
tensors \cite{Jud67}. Equation (\ref{natoli}), which is restricted to the case of 
full circular polarisation ($P_c=1$), is derived by 
neglecting relativistic corrections 
to the radial part of the atomic wave functions \cite{Tho92,Car93a}.

Our task is to identify the physical observable 
defined by the hermitean one-electron operator 
\begin{displaymath}
i\sum_{m,m'}\left(C^{20}_{lm;l'm'}a^{\dagger}_{lm}\tilde{a}_{l'm'}
- {\rm h.c. }\, \right)\, , 
\end{displaymath}
with $l'=l\pm1$. To this end, 
we define 
\begin{equation}
{\bbox A}={\bbox n}\,f_1(N_0)+{\bbox \nabla}_{\Omega}f_2(N_0)\, ,
\label{gens}
\end{equation}
with ${\bbox n}={\bf r}/r$ and 
${\bbox \nabla}_{\Omega}=-i{\bbox n}\times{\bbox l}$; 
${\bbox l}$ denotes the orbital angular momentum. Also, 
\begin{displaymath}
f_1(N_0)=(N_0 - \frac{1}{2})\sqrt{(N_0-1)/N_0}
\end{displaymath}
and 
\begin{displaymath}
f_2(N_0)=\sqrt{(N_0-1)/N_0},
\end{displaymath}
with 
$N_0|lm\rangle=(l+\frac{1}{2})|lm\rangle$. 

The action of ${\bbox A}$, ${\bbox A}^{\dagger}$, ${\bbox l}$ and $N_0$ 
on the spherical harmonics identifies a representation of $o_{3,2}$, 
a rank-two Lie algebra \cite{Eng72}. The corresponding de Sitter group, 
$O(3,2)$, has been used by Goshen and Lipkin 
to describe rotational and vibrational states of nucleons 
in a 2$d$ harmonic-oscillator potential \cite{GoL59,GoL68}. 
In their work, the $O(3,2)$ generators are represented using 
Schwinger's uncoupled-boson scheme \cite{Bie65}. Our considerations will be based on 
representation (\ref{gens}) from which, we believe, physical properties are easier 
to grasp. [The possibility of mapping our problem onto a two-dimensional 
harmonic oscillator reflects the fact that Eq. (\ref{natoli}) 
depends only on two angular variables.]

The Wigner-Eckart theorem yields 
\begin{eqnarray}
\sum_m C^{20}_{lm;l+1-m}&&a^{\dagger}_{lm}\tilde{a}_{l+1-m}
\\
&&
=c_l\, \sum_m 
\langle l m|\left({\bbox A},{\bbox l}\right)^{(2)}_0|l+1 m\rangle \,a^{\dagger}_{l m} 
a_{l+1 m}\, ,
\nonumber
\end{eqnarray}
with $c_l = -\sqrt{10}/\sqrt{l(2l+1)(l+1)(l+2)(2l+3)}$ \cite{VMK88}. Setting 
\begin{equation}
\sum_n\left[\left({\bbox A},{\bbox l}\right)^{(2)}_{\rho}\right]^{l,l'}_n
= \sum_{mm'} 
\langle l m|\left({\bbox A},{\bbox l}\right)^{(2)}_{\rho}|l' m'\rangle 
\,a^{\dagger}_{l m}a_{l' m'}
\label{pseudo_dev} 
\end{equation}
extends the definition of coupled tensors \cite{Jud67} to 
$l,l'$ pairs. [The couplings are defined by $(U^{(s)},V^{(t)})^{(k)}_{\rho}
=\sum_{\nu\mu}C_{s\nu;t\mu}^{k\rho}U^{(s)}_{\nu}V^{(t)}_{\mu}$, where $s,t$ and 
$k$ denote the ranks of the corresponding irreducible tensors; $s=t=1$ and 
$k=2$ in Eq. (\ref{pseudo_dev}).] 
We thus have
\begin{eqnarray}
i\sum_m &&\left(C^{20}_{lm;l+1-m}a^{\dagger}_{lm}\tilde{a}_{l+1-m} 
- {\rm h.c. }\, \right)
\nonumber
\\
&&= i\, c_l\, \sum_n \left[ ({\bbox A},{\bbox l})^{(2)}_0
-\,  ({\bbox l},{\bbox A}^{\dagger})^{(2)}_0 \right]^{l,l+1}_n,
\label{coup_tens}
\end{eqnarray}
where the relation,
\begin{displaymath}
\left[({\bbox A},{\bbox l})^{(x)}_{\rho}\right]^{\dagger}
=(-1)^{x-\rho}({\bbox l},{\bbox A}^{\dagger})^{(x)}_{-\rho}\, ,
\end{displaymath}
has been used. 
A similar result is obtained for the symmetry related pair $l,l-1$. 
The rank-two tensor given by Eq. (\ref{coup_tens}) is even under time reversal 
and odd under space inversion. Tensors with these symmetry properties are 
known as pseudodeviators. (We recall that the expectation value of inversion-odd 
operators vanishes in any state of definite parity.) 

Writing out the the totally symmetric components, {\it e.g}
$
({\bbox A},{\bbox l})^{(2)}_0= \left[ 3A_0 l_0 - 
{\bbox A}\cdot{\bbox l}\right]/\sqrt{6}\, ,
$
and using the orthogonality relations 
${\bbox A}\cdot{\bbox l}={\bbox l}\cdot{\bbox A}^{\dagger}=0$, 
the r.h.s. of Eq. (\ref{coup_tens}) can be given the simpler form
\begin{equation}
i\, c_l \sqrt{\frac{3}{2}}\, \sum_n \left[(A_0  - 
A^{\dagger}_0)l_0 \right]^{l,l+1}_n\,.
\end{equation}
The symmetry group is thus restricted to the subgroup 
$O(3,2)\subset O(2,1)\times O(2)$. [$O(2,1)$ is a three dimensional 
Lorentz group with generators $A_0, A_0^{\dagger}$ 
and $N_0$; $l_0$ commutes with all of them \cite{Bie65}.] 

A sum rule for XNCD, integrated over the two partners $j_{\pm}$, 
can now be written. It reads
\begin{eqnarray}
&&
\Sigma_{\rm XNCD} =
\frac{2\pi^2\alpha}{\hbar c}
(3\cos\theta^2 - 1)
(2c+1)
\label{E1-E2}
\\
&&
\sum_{l=c\pm1 \atop l'=l\pm1}
R_{cl}^{(1)}R_{cl'}^{(2)}a_{l'}(c,l)
\langle g|\sum_n i\left[(A^{\dagger}_0  - A_0)l_0 \right]^{l,l'}_n
| g \rangle 
\nonumber
\end{eqnarray}
where $\cos\theta = \hat{\bf k}\cdot\hat{\bf z}$, with $\hat{\bf z}$ 
the quantisation axis, and
\begin{displaymath}
a_{l+1}(c,l) = \frac{\sqrt{(2l+1)(2l+3)}
[4+3c(c+1)-l(3l+5)]}{(c-l-3)
(c+l+4)(c+l)^2(c+l+2)^2}\, ,
\end{displaymath}
\begin{displaymath}
a_{l-1}(c,l) =\frac{\sqrt{(2l+1)(2l-1)}[6+3c(c+1)-l(3l+1)]}{(c-l+3)
(c+l-2)(c+l)^2(c+l+2)^2}\, . 
\end{displaymath}
Equation (\ref{E1-E2}) is consistent with Kuhn's natural dichroism 
sum rule \cite{Smi76}.

As x-ray linear dichroism experiments in non-centro-\linebreak symmetric 
crystals are currently under work, it seems appropriate to extend our 
analysis of E1-E2 integrated spectra to the case of arbitrary polarisation. 
We obtain
\begin{eqnarray}
\int_{j_++j_-} \frac{\sigma^{\bbox \epsilon}_{\rm X}(\omega)}
{(\hbar\omega)^2} d(\hbar\omega)&&\propto \sum_{x=1}^3 
\left\{
\begin{array}{ccc}
l & l' & x \\
2 & 1 & c\\
\end{array}
\right\}
T^{(x)}_0({\bbox \epsilon},{\bf k})
\label{arb_pol}
\\
&&
i\sum_{mm'}\langle g | C^{x0}_{lm;l'm'}
a^{\dagger}_{lm}\tilde{a}_{l'm'}
- {\rm h.c.}\,|g\rangle\, ,
\nonumber
\end{eqnarray}
with the polarisation response  given by
\begin{displaymath}
T^{(x)}_0({\bbox \epsilon},{\bf k})
=\sum_{\alpha\beta\delta\zeta}
C^{x0}_{1\delta;2\zeta} Y_{1\delta}({\bbox \epsilon}^*)
C^{2\zeta}_{1\alpha;1\beta}Y_{1\alpha}({\bbox \epsilon})
Y_{1\beta}({\bf k})\, . 
\end{displaymath}
As observed, circular dichroism picks out the
$x=2$ term. The remaining values, $x=1,3$, 
are selected by linear dichroism. These contributions are associated with 
time-odd electronic properties of non-centrosymmetric crystals.
For $x=1$, we find
\begin{eqnarray}
&&
T^{(1)}_0({\bbox \epsilon},{\bf k})\,i\sum_{mm'}
\left( 
C^{10}_{lm;l'm'} a^{\dagger}_{lm}\tilde{a}_{l'm'}
- {\rm h.c.}\right)
\nonumber
\\
&&
\propto \hat{\bf k}\cdot\left( {\bbox A}\times {\bbox l}
-{\bbox l} \times{\bbox A}^{\dagger}\right)\, ,
\label{vector} 
\end{eqnarray}
providing a microscopic expression of the irreducible vector operator.  
A physical interpretation of the results (\ref{coup_tens}) and 
(\ref{vector}) is provided below.

For simplicity, consider a single ion with a partially filled valence 
shell in configuration $l^n$; all other shells filled. The effect of 
spin-orbit interactions and/or crystal fields results in a deformation 
of the electronic cloud. Its multipolar expansion will contain spin 
and orbital moments characteristic of 
the symmetry of the deformation and described by one-particle coupled 
tensors. As shown in previous work \cite{Tho92,Car93a,Car93b}, linear 
and circular x-ray dichroism, from pure electric-multipole transitions, 
provide a measure of the ground-state expectation value of these tensors. 
As is well known, integrating over the two partners of a spin-orbit split 
inner shell singles out orbital moments, which can all be constructed 
by coupling the spherical components of ${\bbox l}$.

Inclusion of hybridisation amounts to considering a valence shell given 
as a superposition of states with different $l$ values. 
New electronic moments, stemming from $l,l'$ pairs of states and 
probed by electric-multipole interferences, appear in this case. We
will refer to them as {\it intrinsic hybridisation moments}. They are 
expressible by way of coupled tensors and the current work has shown how 
to write them out in terms of de Sitter generators, for the spinless case.
The E1-E2 interference is sensitive to space-odd tensors. (The orbital 
pseudodeviator represents a space-odd intrinsic hybridisation moment of rank 2.) 
Space-even moments could be revealed by the E1-E3 interference, if ever 
observable.

The use of Eq. (\ref{gens}) 
leads to expressions for the hybridisation moments 
in terms of position and momentum operators, with ${\bf r}\rightarrow
{\bbox n}$ and $i{\bf p} = {\bbox n}
\frac{\partial}{\partial r} +\frac{1}{r}{\bbox \nabla}_{\Omega}
\rightarrow {\bbox \nabla}_{\Omega}$, since we are dealing with angular 
variables only.

{\em E1-M1 interference}. In the optical range (O), 
the absorption cross section is controlled by the E1-M1 interference 
and reads
\begin{eqnarray}
&&
\sigma^{\bbox \epsilon}_{\rm O}(\omega) = 4\pi^2\alpha\hbar\omega
\left[
\frac{i}{2}\sum_f 
\sum_{nn'}
\langle g\mid {\bbox \epsilon}^{\ast}\cdot {\bf r}_n
\mid f\rangle
\label{cross_section2}
\right.
\\
&&
\left.
\phantom{\sum_f}
\langle f\mid {\bbox \epsilon} \times 
{\bf k}\cdot {\bbox l}_{n'} \mid g\rangle 
+ {\rm c.c.}
\right]
\delta(E_f-E_g-\hbar\omega)\, .
\nonumber
\end{eqnarray}
Proceeding as in the derivation of Eq. (\ref{natoli}), effective operators 
of rank zero and two are obtained by coupling the generators of $O(3,2)$. 
Notice that two-particle operators are found in this case, as we are 
considering intra-shell transitions. (Details of their derivation 
will be given elsewhere.) For the rank-zero tensor, we find
\begin{equation}
\sum_{n\neq n'} \left[\,i \left({\bbox A}_{n}-
{\bbox A}_{n}^{\dagger}\right)\cdot {\bbox l}_{n'}
\right]^{l,l'}\, ,
\label{pseudoscal}
\end{equation}
with $l'=l\pm1$. 
(Its rank-two companion, {\it i.e}. the 
two-particle pseudodeviator will not be discussed.) 
Expression (\ref{pseudoscal}) defines an {\it orbital pseudoscalar}, an 
irreducible tensor able to distinguish between enantiomeric and 
non-enantiomeric systems. Indeed,  
it does not branch to the totally symmetric representation in the 
allowed point groups which contain a mirror plane. 
(These branchings can be obtained from Butler's tables \cite{But80}; 
see also Table I in Jerphagnon and Chemla \cite{Jer76}.) 

In the O(3,2) framework an exhaustive picture of one-electron effects 
accessible to x-ray dichroism is obtained. We distinguish four cases:
\newline 
1. Time-odd electronic properties in centrosymmetric crystals. 
They are detected by XMCD. For E1 transitions, the 
spinless effective operator coincides with $l_0$ and we recover the 
familiar orbital sum rule \cite{Tho92}.\newline
2.  Time-even electronic properties in centrosymmetric crystals. 
They are detected by x-ray linear dichroism. For E1 transitions, the 
spinless effective operator coincides with $3l_0^2-{\bbox l}^2$ and the 
orbital-quadrupole sum rule is obtained \cite{Car93b}.\newline
3. Time-even electronic properties in non-centrosym-\linebreak metric crystals. 
They are detected by x-ray circular dichroism. For E1-E2 interference, the spinless 
effective operator is given by Eq. (\ref{coup_tens}), and we have the 
orbital-pseudodeviator sum rule. (If ferromagnetism is present, 
pure electric multipole transitions will also contribute yielding time-odd 
orbital tensors, usually detected by XMCD.  
These terms vanish when the magnetisation direction is perpendicular 
to the photon wave vector.)\newline
4. Time-odd electronic properties in non-centrosym-\linebreak metric crystals. 
They are detected by x-ray linear dichroism. For E1-E2 interference, two 
spinless effective operators contribute as shown by Eq. (\ref{arb_pol}). 
(In the case of a magnetic crystal, pure electric transitions will also
contribute yielding time-even tensors. These terms can be  distinguished 
by full angular-dependence analysis.)

We note that integrating over a single partner ($j_{\pm}$) 
would provide spin-dependent intrinsic hybridisation moments.
The generalisation of our results to x-ray resonant scattering 
would also be straightforward. 

To summarise: We have discussed a theory for x-ray dichroism which is 
applicable to both centro- and non-centrosymmetric crystals. Our 
formalism is constructed from the generators of the O(3,2) group.
Previous theoretical work on integrated dichroic spectra 
\cite{Tho92,Car93a,Car93b} now appears as special case. 

Stimulating discussions with F. de Bergevin, M. Fabrizio, J. Goulon, 
B. R. Judd, C. R. Natoli, F. Pistolesi, and A. Rogalev are gratefully 
aknowledged.

\begin{thebibliography}{99}
%
\bibitem[\star]{AAuth} Present address: Institut f\"{u}r Theoretische 
Physik C, Technische Hochschule, 52074 Aachen, Germany.
\bibitem{Sch87} G. Sch\"{u}tz, W. Wagner, W. Wilhelm, P. Kienle, 
 R. Zeller, R. Frahm, and G. Materlik, 
 Phys. Rev. Lett. {\bf 58}, 737 (1987).
\bibitem{Che90} C. T. Chen, F. Sette, Y. Ma, and S. Modesti, 
 Phys. Rev. B {\bf 41}, 9766 (1990).
\bibitem{Car91} P. Carra, B. N. Harmon, B. T. Thole, M. Altarelli,
 and G. A. Sawatzky, Phys. Rev. Lett. {\bf 66}, 2495 (1991). 
\bibitem{Lan95} J. C. Lang, G. Srajer, C. Detlefs, A. I. Goldman, 
H. K\"{o}nig, X. D. Wand, B. N. Harmon, and R. W. McCallum, 
Phys. Rev. Lett. {\bf 74}, 4935 (1995).
\bibitem{Tho92}  B. T. Thole, P. Carra, F. Sette, and G. van der Laan, 
 Phys. Rev. Lett. {\bf 68}, 1943 (1992). 
\bibitem{Car93a}  P. Carra, B. T. Thole, M. Altarelli, and X. D. Wang, 
 Phys. Rev. Lett. {\bf 70}, 694 (1993).
\bibitem{Car93b}  P. Carra, H. K\"{o}nig, B. T. Thole, and M. Altarelli, 
 Physica B {\bf 192}, 182 (1993). 
\bibitem{BCA00} R. Benoist, P. Carra, and O. K. Andersen, submitted for publication.
\bibitem{Gou98} Goulon J., C. Goulon-Ginet, A. Rogalev, V. Gotte, C. Malgrange, 
 C. Brouder, and C. R. Natoli, J. Chem. Phys. {\bf 108}, 6394 (1998).
\bibitem{Ala98} Alagna L., T. Prosperi, S. Turchini, J. Goulon, 
 A. Rogalev, C. Goulon-Ginet, C. R. Natoli, R. D. Peacock, 
 and B. Stewart, Phys. Rev. Lett. {\bf 80}, 4799 (1998).
\bibitem{Nat98} C. R. Natoli, Ch. Brouder, Ph. Sanctavict, J. Goulon, 
 Ch. Goulon-Ginet, and A. Rogalev, Eur. Phys. J. B {\bf 4}, 1 (1998). 
\bibitem{Jud67} B. R. Judd, {\em Second Quantization and Atomic Spectroscopy}, 
 The Johns Hopkins Press, Baltimore, 1967.
\bibitem{Eng72} M. J. Englefield, {\it Group Theory and the Coulomb Problem}, 
 Wiley-Interscience, New York, 1972.
\bibitem{GoL59} S. Goshen and H. J. Lipkin, Ann. Phys. (N. Y.) {\bf 6}, 301 (1959).
\bibitem{GoL68} S. Goshen and H. J. Lipkin, in {\it Spectroscopic and Group Theoretical 
 Methods in Physics - Racah Memorial Volume}, F. Bloch, S. G. Cohen, A. De-Shalit, 
 S. Sambursky, and I. Talmi Eds., North-Holland, Amsterdam, 1968.
\bibitem{Bie65} J. Schwinger, in {\it Quantum Theory of Angular Momentum}, L. C. 
 Biedenharn and H. van Dam Eds., Academic Press, New York, 1965.
\bibitem{VMK88} The matrix elements of $({\bbox n},{\bbox l})^{(2)}$ 
and $({\bbox \nabla}_{\Omega},{\bbox l})^{(2)}$ are given by D. A. Varshalovich, 
 A. N. Moskalev, and V. K. Khersonskii, {\it Quantum Theory of Angular Momentum} 
 (World Scientific Publishing, Singapore, 1988), Chap. 13, p. 493.
\bibitem{Smi76} D. Y. Smith, Phys. Rev. {\bf 13}, 5303 (1976).
\bibitem{But80} P. H. Butler, {\it Point Group Symmetry Applications - Methods and 
 Tables}, Plenum Press, New York, 1980. 
\bibitem{Jer76} J. Jerphagnon and D. S. Chemla, J. Chem. Phys. {\bf 65}, 1522 (1976).
%
\end{thebibliography}
\end{document}